\documentclass[prb,twocolumn,showpacs]{revtex4}

\usepackage{graphicx,epsfig}
\usepackage{bm}

\begin{document}

\title{Adiabatic pumping through a quantum dot in the Kondo
       regime:\\
       Exact results at the Toulouse limit}
\smallskip

\author{Avraham Schiller$^{1}$ and Alessandro Silva$^{2}$}

\affiliation{
             $^1$ Racah Institute of Physics, The Hebrew
                  University, Jerusalem 91904, Israel\\
             $^2$ Abdus Salam ICTP, Strada Costiera 11,
                  34100 Trieste, Italy}

\begin{abstract}
Transport properties of ultrasmall quantum dots with a single
unpaired electron are commonly modeled by the nonequilibrium Kondo
model, describing the exchange interaction of a spin-$1/2$ local
moment with two leads of noninteracting electrons. Remarkably, the
model possesses an exact solution when tuned to a special manifold
in its parameter space known as the Toulouse limit. We use the
Toulouse limit to exactly calculate the adiabatically pumped spin
current in the Kondo regime. In the absence of both potential
scattering and a voltage bias, the instantaneous charge current is
strictly zero for a generic Kondo model. However, a nonzero spin
current can be pumped through the system in the presence of a finite
magnetic field, provided the spin couples asymmetrically to the two
leads. Tunneling through a Kondo impurity thus offers a natural
mechanism for generating a pure spin current. We show, in
particular, that one can devise pumping cycles along which the
average spin pumped per cycle is closely equal to $\hbar$. By
analogy with Brouwer's formula for noninteracting systems with two
driven parameters, the pumped spin current is expressed as a
geometrical property of a scattering matrix. However, the relevant
scattering matrix that enters the formulation pertains to the
Majorana fermions that appear at the Toulouse limit rather than the
physical electrons that carry the current. These results are
obtained by combining the nonequilibrium Keldysh Green function
technique with a systematic gradient expansion, explicitly exposing
the small parameter controlling the adiabatic limit.
\end{abstract}


\maketitle

\section{Introduction}
\label{sec:Introduction}

The act of pumping is well known from everyday life.
By repeatedly operating a periodic sequence of steps
one can transfer a certain amount of fluid or gas
between reservoirs held at equal potential. The same
principle applies to electrical charge. By periodically
modulating spatially-confined potentials it is possible
to generate a nonzero dc current between leads that are
kept at equal temperature and electrochemical potential.
When operated sufficiently slow such that the typical
scattering time for electrons is much faster than the
time over which the scattering potentials vary, this
process is know as adiabatic quantum pumping. Recently,
there has been considerable theoretical~\cite{Thouless83,
Brouwer98,AA98,ZSA99,SAA00,AEGS01,VAA01,MM01,EAL02,MB02,
WW02,SC01-03,CAN03,Aono04,MCM02,GTF03,Zheng03,BF05,SB03}
and experimental~\cite{Kouwenhoven91,Pothier92,Switkes99,
Watson99} interest in adiabatic quantum pumping in
confined nanostructures. Besides the fundamental and
technological importance of understanding time-dependent
phenomena in nano-devices such as semiconductor and
carbon nanotube quantum dots, adiabatic quantum pumping
offers new possibilities that otherwise are difficult
to realize in conventional dc transport measurements
with a finite voltage bias. Most notably, the
ability to pump a quantized amount of charge per
cycle,~\cite{Kouwenhoven91,Switkes99} which is of
potential metrological importance. In this paper we
address another such example, the generation of pure,
possibly quantized spin current without any charge
current.~\cite{Aono04,SC01-03,MCM02,GTF03,Zheng03,BF05,
SB03,Watson99}

In the absence of interactions, adiabatic pumping is by now
well understood. In particular, building on the scattering
approach of B\"uttiker {\em et al.},~\cite{BTP94} Brouwer
has elegantly shown~\cite{Brouwer98} that the adiabatically
pumped current can be expressed in terms of the
instantaneous (equilibrium) scattering matrix. In the
case of two driven parameters, the pumped charge per cycle
reduces to a geometrical property of the equilibrium
scattering matrix, pertaining to the area enclosed in
parameter space by the pumping cycle. All other details
of the pumping cycle, i.e., the explicit time dependences
of the scattering potentials, are irrelevant as long as
pumping is adiabatic.

Far less understood are the effects of interactions,
where efforts have focused thus far on
zero-~\cite{AA98,WW02,Aono04} and
one-dimensional~\cite{SC01-03,CAN03} systems. The
difficulty with incorporating interactions lies in the
need to treat retardation effects beyond the static limit.
Indeed, recent attempts to generalize Brouwer's formula
so as to include interactions~\cite{SGKF05,SO05,Fiore07}
have either required the introduction of complicated
vertex corrections,~\cite{SGKF05,SO05} or the application
of a gradient expansion to interaction-induced
self-energies.~\cite{Fiore07} Both formulations
can only be implemented approximately at this stage,
urging the need for benchmark results against which
approximate treatments can be tested. In this paper
we provide such an exact result for the pumped
currents through a Kondo impurity.

Kondo-assisted tunneling has been observed by now in an abundance of
nanostructures, ranging from semiconductor~\cite{sc-QD} and
nanotube~\cite{NT-QD} quantum dots, to
single-atom~\cite{single-atom} and
single-molecule~\cite{single-molecule} transistors. In the Kondo
regime, these systems are described by the well-known Kondo model: a
spin-$\frac{1}{2}$ local moment undergoing antiferromagnetic
spin-exchange with the conduction electrons in the leads. The
nonequilibrium Kondo model, either with a static or a time-dependent
voltage bias, is a difficult problem. Remarkably, it possesses an
exact solution when tuned to a special manifold in its parameter
space, known as the Toulouse limit.~\cite{SH95,SH98} At the Toulouse
limit one can apply a suitable canonical transformation to recast
the interacting problem in free, quadratic form. This requires the
introduction of new fermionic degrees of freedom having no simple
relation to the physical electrons in the leads. The resulting
solution, which generalizes previous exact results for the
equilibrium Kondo problem,~\cite{Toulouse70,EK92} does not
correspond to realistic parameters. It requires large values of
certain exchange couplings (see below), rendering it incapable of
describing weak-coupling physics. However, the Toulouse limit is
expected to correctly describe the strong-coupling regime of the
nonequilibrium Kondo effect, as different microscopic models are
governed by the same strong-coupling fixed point. Indeed, previous
applications of the model to dc,~\cite{SH95,SH98} ac,~\cite{SH96}
and pulsed-bias potentials~\cite{SH00} have shown all the
qualitative features of Kondo-assisted tunneling: a zero-bias
anomaly that splits in an applied magnetic field; Fermi-liquid
characteristics in the low-$T$ and low-$V$ differential conductance;
side peaks in the differential conductance at $eV = \pm n\hbar
\omega$ for an ac drive of frequency $\omega$; and a hierarchy of
time scales for the rise, saturation and falloff of the current in
response to a pulsed bias potential. The Toulouse limit was also
recently applied to compute the full counting statistics for
tunneling through a Kondo impurity.~\cite{KG05,SKG06}

In this paper we take the solution one step further,
by exactly computing the adiabatically pumped currents
on the Toulouse manifold. Contrary to previous
applications of the Toulouse limit to Kondo-assisted
tunneling we set the voltage bias to zero, but consider
a general periodic modulation of the transverse
exchange couplings and the local magnetic field (the
free parameters on the Toulouse manifold). In the limit
of slow time variations we obtain an exact analytic
expression for the adiabatically pumped spin current.
In particular, we show that a nonzero spin current
can be pumped through the system for a time-varying
magnetic field, provided the couplings to the two
leads are made asymmetric. Such a condition is
easily met in practical devices. Unlike the spin
current, the instantaneous charge current is strictly
zero in the absence of both potential scattering and
a voltage bias, as follows from general symmetry
considerations. This feature is generic to the Kondo
model, independent of the adiabatic and Toulouse
limits. Hence tunneling through a Kondo impurity
offers a natural mechanism for the realization of
a spin battery, i.e., a source of pure spin current
without any charge current. This statement, valid
both in the adiabatic limit and beyond, is in
qualitative agreement with earlier slave-boson
mean-field studies of adiabatic pumping through an
Anderson impurity,~\cite{Aono04,SGKF05} indicating
that no fine tuning of model parameters is required
as long as one operates in the Kondo regime. Finally,
we show that one can devise suitable pumping cycles
that operate as a quantized spin pump. Namely, a
spin closely quantized to $\hbar$ is pumped per
cycle without an accompanying charge.

As indicated above, the solution at the Toulouse
limit relies on a nonlocal transformation that
converts the original spin-exchange Hamiltonian to
free-fermion form.~\cite{EK92,SH98} In contrast to
conventional quadratic Hamiltonians, though, the number
of fermions (not to be confused with the physical
electrons in the system) is not conserved, excluding the
application of Brouwer's formula in its existing form.
To generalize Brouwer's result to this somewhat
unconventional case, we follow a path similar to
the one taken by Vavilov {\em et al.}~\cite{VAA01}
in studying the photovoltaic effect in open chaotic
cavities. Starting from the nonequilibrium Keldysh
Green function technique, we show how the adiabatic
limit is obtained from a systematic gradient
expansion. In this manner we are able to express
the instantaneous spin current in terms of an
energy-shift matrix,~\cite{AEGS01} leading to a
Brouwer-type formula for the adiabatically pumped
spin current.

The formalism outlined above has three notable
advantages over the scattering approach~\cite{BTP94}
originally used by Brouwer to derive his result:
(i) It conveniently accommodates the case where
    particles are not conserved;
(ii) All orders of perturbation theory are summed
     up in the Keldysh technique, thus exceeding
     linear response;
(iii) Based on a systematic gradient expansion,
     one can easily read off the small parameter
     controlling the adiabatic limit.
We emphasize, however, that the resulting Brouwer-type
formula for the electronic spin current is formally
expressed in terms of the scattering matrix for the Majorana
fermions that appear in the transformed Hamiltonian. While
technically useful, these degrees of freedom have neither a simple
representation nor interpretation in terms of the physical electrons
in the leads, thus obscuring a clear physical picture.
It remains to be seen whether a similar expression can be written
down for the spin current directly in terms of the scattering
properties of the lead electrons which carry the current.

\begin{figure}[tb]
\centerline{
\includegraphics[width=75mm]{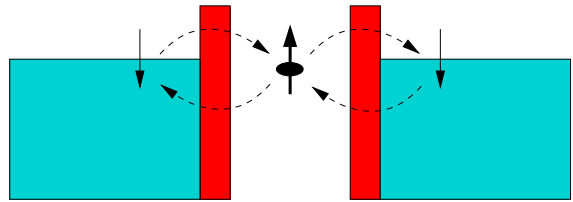}
}\vspace{0pt}
\caption{Schematic description of the physical system. A
         spin-$\frac{1}{2}$ local moment $\vec{\tau}$ is
         placed in between two leads of noninteracting
         spin-$\frac{1}{2}$ electrons. The local moment
         $\vec{\tau}$ experiences a spin-exchange
         interaction with the local conduction-electron
         degrees of freedom near the junction, as described
         by the Hamiltonian of Eq.~(\ref{Full_H}).
         Tunneling between the leads is facilitated by
         spin-exchange terms that scatter an electron
         across the junction.}
\label{Fig:fig1}
\end{figure}

The remainder of the paper is organized as follows.
In Sec.~\ref{sec:Toulouse-limit} we briefly review
the Toulouse limit, introducing the different Green
functions that will later be used in the course of
the calculation. In Sec.~\ref{sec:symmetries} we
present general symmetry considerations and apply
them to the problem at hand. In particular, we show
that the instantaneous charge current is strictly
zero in the absence of potential scattering, whereas
the spin current is zero unless the dot couples
asymmetrically to the two leads. Proceeding with
quantitative calculations, we combine the Keldysh
technique with a gradient expansion in
Sec.~\ref{sec:spin-current} to derive a Brouwer-type
formula for the adiabatically pumped spin current in
the Toulouse limit. Using this formula, a specific
class of pumping cycles is analyzed in detail in
Sec.~\ref{sec:results}. In particular, we demonstrate
a pumping cycle for which the total spin pumped per
cycle is closely equal to $\hbar$, thus operating
as a quantized spin pump. Finally, we present our
conclusions in Sec.~\ref{sec:discussion}.

\section{Physical model and Toulouse limit}
\label{sec:Toulouse-limit}

We begin with a brief review of the Toulouse limit,
and with introducing the different Green functions
that will later be used in calculating the pumped
spin current. The physical system under consideration
is shown schematically in Fig.~\ref{Fig:fig1}. A
spin-$\frac{1}{2}$ local moment $\vec{\tau}$ is
embedded between two leads of noninteracting
spin-$\frac{1}{2}$ electrons, undergoing a
spin-exchange interaction with the local
conduction-electron degrees of freedom on either side
of the junction. As emphasized in the introduction,
the impurity moment $\vec{\tau}$ can either represent
an ultrasmall quantum dot with a single unpaired
electron,~\cite{sc-QD} or an actual magnetic
impurity as in single-atom~\cite{single-atom} and
single-molecule~\cite{single-molecule} transistors.

Since scattering off the impurity is restricted to the
$s$-wave channel, one can reduce the conduction-electron
degrees of freedom that couple to the impurity to
one-dimensional fields $\psi_{\alpha \sigma}(x)$, where
$\alpha = R, L$ labels the lead (right or left) and
$\sigma = \uparrow, \downarrow$ specifies the spin
orientations. In terms of the one-dimensional fields,
coupling to the impurity takes place via the local spin
densities at the origin:
$\vec{s}_{\alpha \beta} = \frac{1}{2}
\sum_{\sigma,\sigma'} \psi^{\dagger}_{\alpha \sigma}(0)
\vec{\sigma}_{\sigma, \sigma'} \psi_{\beta \sigma'}(0)$.
The most general form of a spin-exchange Hamiltonian is
therefore
\begin{eqnarray}
{\cal H} &=& i v_{F}\sum_{\alpha = L, R}\,
               \sum_{\sigma = \uparrow,\downarrow}
               \int_{-\infty}^{\infty}\!
               \psi_{\alpha \sigma}^{\dagger}(x)
               \partial_{x} \psi_{\alpha \sigma}(x) dx
\nonumber \\
          &+& \!\! \sum_{\alpha, \beta = L, R}\,
              \sum_{\lambda = x, y, z}
                 J_{\lambda}^{\alpha \beta}(t)
                 \tau^{\lambda} s^{\lambda}_{\alpha \beta}
            - \mu_B g_i H(t) \tau^z ,
\label{Full_H}
\end{eqnarray}
where we have allowed for different exchange couplings
$J_{\lambda}^{\alpha \beta} = J_{\lambda}^{\beta \alpha}$,
and for a local magnetic field $H$ acting on the impurity
spin. Here $\mu_B$ and $g_i$ are the Bohr magneton and
impurity Land\'e $g$ factor, respectively. Throughout
the paper we use units for which $\hbar = k_B = 1$, while
the electronic charge is taken to be $-e$. Proper units
will be reinstated in some of the final expressions
presented below.

The Hamiltonian of Eq.~(\ref{Full_H}) is written
for general time-dependent exchange couplings
$J_{\lambda}^{\alpha \beta}(t)$ and local magnetic
field $H(t)$. Our interest, however, will be in
slow periodic modulations of the transverse couplings
$J_{x}^{\alpha \beta}(t) = J_{y}^{\alpha \beta}(t)$
and the local magnetic field. The longitudinal
couplings $J_{z}^{\alpha \beta}$ will be taken
to be constant in time and equal to particular
values as detailed below. It is this fine tuning
of $J_{z}^{\alpha \beta}$ that defines the Toulouse
manifold and which enables our exact solution.

\subsection{Toulouse limit}

The spin-exchange Hamiltonian of Eq.~(\ref{Full_H})
is conventionally derived from the more basic
Anderson impurity model via the Schrieffer-Wolff
transformation.~\cite{SW66} The couplings
$J_{\lambda}^{\alpha \beta}$ generated in this case
are weak, isotropic (i.e., independent of $\lambda$),
and satisfy $J^{LL} J^{RR} = (J^{LR})^2$. The Toulouse
limit corresponds to a different sector in the parameter
space of the Kondo Hamiltonian where $J_z^{LL} = J_z^{RR}
= 2\pi v_F$ and $J_z^{LR} = 0$. The transverse couplings
$J_x^{\alpha \beta}(t) = J_y^{\alpha \beta}(t) \equiv
J_{\perp}^{\alpha \beta}(t)$ and the local magnetic
field $H(t)$ are allowed to be arbitrary, and will
subsequently be taken to be periodically modulated
in time. Physically, this choice of parameters
implies that tunneling is always accompanied by a
spin flip. Although quite remote from the situation
encountered in real quantum dots, this model is
expected to correctly describe the strong-coupling
regime of the Kondo effect, as argued in
the introduction and elaborated on in
Refs.~\onlinecite{SH95,SH98} and \onlinecite{SH00}.
In particular, it has been shown~\cite{SH98} that the
strong-coupling physics of the Anderson impurity model
is best described both in and out of equilibrium by
couplings that satisfy
\begin{eqnarray}
J_{\perp}^{LL} J_{\perp}^{RR} = (J_{\perp}^{LR})^2 .
\label{condi}
\end{eqnarray}

As described in detail in Ref.~\onlinecite{SH98},
the Hamiltonian of Eq.~(\ref{Full_H}) can be mapped
under the conditions listed above onto a free-fermion
form. The mapping involves a sequence of steps,
comprised of (i) bosonizing the fermion fields,
(ii) a nonlocal canonical transformation involving
the conduction-electron spin degrees of freedom,
and (iii) refermionization of the boson fields to
form four new fermion fields: $\psi_{\nu}(x)$ with
$\nu = c, s, f, sf$. Here $c, s, f$, and $sf$ stand
for charge, spin, flavor (left minus right), and
spin-flavor fields. In addition, the impurity spin
$\vec{\tau}$, which has been mixed by the canonical
transformation with the conduction-electron spin degrees
of freedom, is represented in terms of two real Majorana
fermions: $\hat{a} = -\sqrt{2}\tau^y$ and $\hat{b} =
-\sqrt{2}\tau^x$. At the conclusion of these steps
one arrives at a quadratic Hamiltonian conveniently
written in the form
\begin{eqnarray}
H' &=&\!\! \sum_{\nu = c, s, f, sf} \sum_k \epsilon_k
             \psi^{\dagger}_{\nu,k}\psi_{\nu,k}
             + i \mu_B g_i H(t)\, \hat{b} \hat{a}
\nonumber \\
   &&\!\!
     +\, i J^{+}_{sf}(t)\, \hat{\chi}^{+}_{sf} \hat{b}
     + i J^{-}_{sf}(t)\, \hat{\chi}^{-}_{sf} \hat{a}
     + i J^{-}_{f}(t)\, \hat{\chi}^{-}_{f} \hat{a} ,
\label{H'}
\end{eqnarray}
where we have introduced the three couplings
\begin{eqnarray}
J^{+}_{sf}(t) &=& \frac{J^{LL}_{\perp}(t)
                        + J^{RR}_{\perp}(t)}
                         {2\sqrt{ 2\pi a}} ,\\
J^{-}_{sf}(t) &=& \frac{J^{LL}_{\perp}(t)
                        - J^{RR}_{\perp}(t)}
                         {2\sqrt{ 2\pi a}} ,\\
J^{-}_{f}(t) &=& \frac{J^{LR}_{\perp}(t)}
                      {\sqrt{ 2\pi a}} .
\end{eqnarray}
Here the energies $\epsilon_k$ are equal to $-v_F k$,
$a$ is an ultraviolet momentum cutoff corresponding
to a lattice spacing, and $L$ is the effective size
of the leads (i.e., $k$ is discretized in units of
$2\pi/L$). The fields $\hat{\chi}^{\pm}_{\nu}$ ($\nu
= f, sf$) are local Majorana fermions, defined as
\begin{eqnarray}
\hat{\chi}^{+}_{\nu} &=& \frac{1}{\sqrt{2 L}}
     \sum_k (
              \psi^{\dagger}_{\nu, k} + \psi_{\nu, k}
            ) ,
\label{Local-Majoranas+} \\
\hat{\chi}^{-}_{\nu} &=& \frac{1}{i\sqrt{2 L}}
     \sum_k (
              \psi^{\dagger}_{\nu, k} - \psi_{\nu, k}
            ) .
\label{Local-Majoranas-}
\end{eqnarray}
Relaxation of each of the conditions $J_z^{LR} = 0$,
$J_z^{LL} + J_z^{RR} = 4\pi v_F$, and $J_z^{LL} -
J_z^{RR} = 0$ introduces a different interaction term
into the Hamiltonian of Eq.~(\ref{H'}), as discussed
in Ref.~\onlinecite{MSH98} and detailed below.

Although noninteracting, the Hamiltonian of Eq.~(\ref{H'})
is unconventional in the sense that it does not conserve
the number of $\psi$ fermions (not to be confused with
the physical electrons in the system). Indeed, the
fermion fields $\psi_{\nu}(x)$ with $\nu = c, s, f, sf$
have neither a simple representation nor a simple
interpretation in terms of the original electronic
degrees of freedom. Consequently, not all observables
can be computed based on the mapping of Eq.~(\ref{H'}).
Only observables that have a simple representation
in terms of the $\psi$ fields are accessible.
Fortunately, both the charge and spin currents
fall in this category.

To derive the transformed forms of the electronic charge
and spin currents, it is necessary to go back to their
original representation in terms of the physical
electrons in the leads. Denoting the total number
operator for electrons with spin projection $\sigma$
in lead $\alpha$ by $\hat{N}_{\alpha \sigma}$, the
charge current flowing from right to left is given by
\begin{equation}
\hat{I}_c = -i e [ {\cal H},
                   \hat{N}_{L \uparrow} +
                   \hat{N}_{L \downarrow} ]
          = i e [ {\cal H},
                  \hat{N}_{R \uparrow} +
                  \hat{N}_{R \downarrow} ] .
\label{I_c-e}
\end{equation}
Here ${\cal H}$ is the Kondo Hamiltonian of
Eq.~(\ref{Full_H}). Since charge fluctuations are
excluded on the dot, the instantaneous charge current
outgoing from the left lead (left commutator) is
identical to the instantaneous charge current
flowing into the right lead (right commutator).
This is no longer the case with the spin current,
defined as half the difference in particle currents
between the spin-up and spin-down electrons.
(The factor of one-half comes from the electronic
spin projection in the $z$ direction). Indeed,
the spin currents associated with the left and
right leads differ by a term proportional to
$d\tau^z/dt$, which stems from conservation
of the total spin projection $S^z_{\rm total}$
of the entire system. Fortunately, this
difference in currents has no significance for
our purposes, since $d\tau^z/dt$ averages to zero
over a single pumping cycle. This grants us the
freedom to work with our operator of choice.
In the following we shall concentrate on the
symmetrized spin current, i.e., the average of
the spin currents to the left and to the right
of the impurity, which turns out to be the most
convenient current combination to work with.
With this convention, the (symmetrized) spin
current flowing from left to right is written as
\begin{equation}
\hat{I}_s = \frac{i}{4}
            [ {\cal H},
               \hat{N}_{R \uparrow}
               - \hat{N}_{R \downarrow}
               - \hat{N}_{L \uparrow}
               + \hat{N}_{L \downarrow} ] .
\label{I_s-e}
\end{equation}

Equations~(\ref{I_c-e}) and (\ref{I_s-e}) specify
the electronic charge and spin currents in terms of
the physical electrons. The transformed operators,
$\hat{I}'_c$ and $\hat{I}'_s$, are obtained by
repeating the same sequence of steps as applied to
the Hamiltonian, namely, bosonization, a nonlocal
canonical transformation, and refermionization.
Skipping the details of the algebra~\cite{SH98}
we quote here only the end result:
\begin{equation}
\hat{I}'_{c} =  i e J^{-}_{f}(t)\,
                    \hat{\chi}^{+}_{f} \hat{a} ,
\label{I_c}
\end{equation}
and
\begin{equation}
\hat{I}'_{s} = \frac{i}{2}
       \left[
              J^{-}_{sf}(t)\, \hat{\chi}^{+}_{sf} \hat{a}
              - J^{+}_{sf}(t)\, \hat{\chi}^{-}_{sf} \hat{b}
       \right] .
\label{I_s}
\end{equation}
Note that although these expressions are written in
terms of Majorana fermions, they describe the actual
electronic charge and spin currents flowing in the
system. The unconventional forms of the currents
stem from the nonlocal transformation that has been
applied.

\subsection{Keldysh Green functions}

To compute the spin current, we shall make use of
the nonequilibrium Keldysh Green function technique.
The basic ingredients of the theory are the greater,
lesser, retarded, and advanced Majorana Green
functions, defined as~\cite{Langreth76}
\begin{equation}
G^{>}_{\alpha\beta}(t, t') =
      \langle \hat{\alpha}(t) \hat{\beta}(t') \rangle ,
\label{G>_def}
\end{equation}
\begin{equation}
G^{<}_{\alpha\beta}(t, t')  =
      \langle \hat{\beta}(t') \hat{\alpha}(t) \rangle ,
\label{G<_def}
\end{equation}
\begin{equation}
G^{r,a}_{\alpha\beta}(t, t') =
        \mp i \theta (\pm t \mp t')
        \langle
               \{ \hat{\alpha}(t) , \hat{\beta}(t') \}
        \rangle .
\label{G-def}
\end{equation}
Here $\alpha, \beta \in \{ a, b\}$, while the upper
and lower signs in Eq.~(\ref{G-def}) correspond to
the retarded ($r$) and advanced ($a$) Green functions,
respectively. The curly brackets in Eq.~(\ref{G-def})
denote the anticommutator.

In thermal equilibrium, the Majorana Green functions
are easily found by summing all orders of the
perturbation theory in the time-independent couplings
$J^{\pm}_{sf}$, $J^{-}_{f}$, and $H$. Specifically,
switching over to the energy domain and assuming
the wide-band limit one obtains
\begin{eqnarray}
G^{r,a}(\epsilon)&=&
           \frac{1}{(\epsilon \pm i\Gamma_a)
                    (\epsilon \pm i\Gamma_b)
                    - (\mu_B g_i H)^2}
\nonumber \\
\nonumber \\
       &\times & \left[
       \begin{array}{cc}
             \epsilon \pm i\Gamma_b & -i\mu_B g_i H \\ \\
             i\mu_B g_i H  & \epsilon \pm i\Gamma_a
       \end{array}
       \right] .
\label{G^ra}
\end{eqnarray}
Here we have adopted a $2 \times 2$ matrix notation,
with the indices $1$ and $2$ corresponding to $a$ and
$b$, respectively.

Equation~(\ref{G^ra}) features two new energy scales,
\begin{equation}
\Gamma_a = \pi \rho_0
           \left [
                   ( J^{-}_{f} )^2 +
                   ( J^{-}_{sf} )^2 \right ]
\label{Gamma_a}
\end{equation}
and
\begin{equation}
\Gamma_b = \pi \rho_0 ( J^{+}_{sf} )^2 ,
\label{Gamma_b}
\end{equation}
where $\rho_0 = 1/(2\pi v_F)$ is the density of states
per unit length in the leads. These two scales
determine the widths of the various Majorana spectral
functions, and thus play the role of Kondo temperatures
at the Toulouse limit. The conventional single-channel
Kondo effect is best described by the case where
$\Gamma_a = \Gamma_b \equiv T_K$, corresponding to
the condition specified in Eq.~(\ref{condi}). The
equilibrium greater and lesser Green functions are
related in turn to $G^{r, a}(\epsilon)$ through
standard identities:
\begin{equation}
G^{>}_{\alpha \beta}(\epsilon) = i [1 - f(\epsilon)]
      [ G^{r}_{\alpha \beta}(\epsilon)
        - G^{a}_{\alpha \beta}(\epsilon) ] ,
\label{G^>}
\end{equation}
\begin{equation}
G^{<}_{\alpha \beta}(\epsilon) = i f(\epsilon)
      [ G^{r}_{\alpha \beta}(\epsilon)
        - G^{a}_{\alpha \beta}(\epsilon) ] ,
\label{G^<}
\end{equation}
where $f(\epsilon)$ is the Fermi-Dirac distribution
function.

As emphasized above, Eqs.~(\ref{G^ra})--(\ref{G^<})
are restricted to thermal equilibrium. They do not
apply when any of the couplings $J^{\pm}_{sf}$,
$J^{-}_{f}$, and $H$ is time dependent, which is
the case of interest here. Indeed, time-dependent
couplings are generally difficult to treat
analytically even for noninteracting systems. Below
we shall first derive the instantaneous spin current
for a general time-dependent setting, but will
eventually be interested in slow periodic
modulations of the four coupling constants listed
above. In terms of the original spin-exchange
Hamiltonian of Eq.~(\ref{Full_H}), we allow for
general time variation of the couplings
$J_{\perp}^{\alpha \beta}$ and field $H$, but
demand that the longitudinal exchange couplings
$J_z^{\alpha \beta}$ be held fixed at their
Toulouse-limit values. We exclude variations in
the phase of $J_{\perp}^{LR} = \left (J_{\perp}^{RL}
\right )^{\ast}$, as this corresponds to biasing
the system. Accordingly, we take $J_{\perp}^{LR}
= J_{\perp}^{RL}$ to be real throughout the paper.

\subsection{Deviations from the Toulouse limit}

We conclude this section by briefly describing
the modifications that are introduced into the
Hamiltonian and the current operators upon
departure from the Toulouse manifold. As discussed
in Ref.~\onlinecite{MSH98}, the Hamiltonian of
Eq.~(\ref{H'}) is supplemented by three new
interaction terms away from the Toulouse limit:
\begin{equation}
{\cal H}' \to {\cal H}' + {\cal H}_{\rm int}
\end{equation}
with
\begin{eqnarray}
{\cal H}_{\rm int} &=& -J_z^{LR}\;
             \hat{b}\hat{a}\;
             \hat{\chi}^{-}_{f} \hat{\chi}^{+}_{sf}
         - i \frac{J_z^{+}}{L}
             \hat{b}\hat{a}
             \sum_{k,k'}
             :\!\psi_{s,k}^{\dagger}\psi_{s,k'}\!:
\nonumber \\
         &&-\, i \frac{J_z^{-}}{L}
             \hat{b}\hat{a}
             \sum_{k,k'}
             :\!\psi^{\dagger}_{sf,k}\psi_{sf,k'}\!: .
\label{Hint}
\end{eqnarray}
Here $\hat{\chi}_{\nu}^{\pm}$ are the local
Majorana fields of Eqs.~(\ref{Local-Majoranas+})
and (\ref{Local-Majoranas-}), while
$:\!\psi^{\dagger}_{\nu,k}\psi_{\nu,k'}\!\!:$ stands
for normal ordering with respect to the unperturbed
Fermi sea of the $\psi$ fermions. The three
couplings $J_z^{-} = (J_z^{LL} - J_z^{RR})/2$,
$J_z^{+} = (J_z^{LL} + J_z^{RR})/2 - 2\pi \hbar v_F$
and $J_z^{LR}$ measure the deviations from the Toulouse
manifold in each of the three possible directions in
parameter space. The new tunneling term $J_z^{LR}$
also modifies the current operators $\hat{I}'_c$ and
$\hat{I}'_s$, which take the general forms
\begin{equation}
\hat{I}'_c = i e J^{-}_{f}(t)
                  \hat{\chi}_{f}^{+} \hat{a}
          - e J_z^{LR}(t) \hat{\chi}_{f}^{+}
                  \hat{\chi}_{sf}^{+} \hat{b}\hat{a}
\label{I_c-int}
\end{equation}
and
\begin{equation}
\hat{I}'_{s} = \frac{i}{2}
       \left[
              J^{-}_{sf}(t)\, \hat{\chi}^{+}_{sf} \hat{a}
              - J^{+}_{sf}(t)\, \hat{\chi}^{-}_{sf} \hat{b}
       \right]
       + \frac{J_z^{LR}(t)}{2} \hat{\chi}_{f}^{-}
         \hat{\chi}_{sf}^{-} \hat{b}\hat{a}.
\label{I_s-int}
\end{equation}
Here we have explicitly allowed for time variation
of the new coupling constant $J_z^{LR}$.

\section{Symmetry considerations}
\label{sec:symmetries}

Before proceeding to detailed calculations, in
this section we first present general symmetry
considerations applicable to any two-lead system.
By analyzing their implications for the Kondo
Hamiltonian of Eq.~(\ref{Full_H}), we identify
necessary conditions for finite charge and spin
currents to be pumped through the system.

\subsection{Particle-hole symmetry acting separately
            on each lead}
\label{sub-sec:PH-1}

Consider a general two-lead system where each
lead is represented by a single spinful channel.
The charge current flowing into lead $\alpha$
($\alpha = L, R$) is given by
\begin{equation}
\hat{I}_{c, \alpha} = -i e [ {\cal H},
                     \hat{N}_{\alpha \uparrow} +
                     \hat{N}_{\alpha \downarrow} ] ,
\label{I_c-symmetry}
\end{equation}
while the symmetrized spin current $\hat{I}_s$ flowing
from left to right is specified in Eq.~(\ref{I_s-e}).
Here $\hat{N}_{\alpha \sigma}$ denotes the total
number operator for electrons with spin projection
$\sigma$ on lead $\alpha$. Let us consider
the situation where the time-dependent Hamiltonian,
${\cal H}$, is invariant under a particle-hole
transformation that converts particles on each lead
to opposite-spin holes on the {\em same} lead (i.e.,
$c^{\dagger}_{\alpha,k,\sigma} \to e^{i \varphi_{\alpha\sigma}}
c_{\alpha, -k, \bar{\sigma}}$, where $\bar{\sigma}$
is the spin index opposite to $\sigma$; the phases
$\varphi_{\alpha\sigma}$ are arbitrary). The total
number operator for electrons on lead $\alpha$,
$\hat{N}_{\alpha} \equiv \hat{N}_{\alpha \uparrow}
+ \hat{N}_{\alpha \downarrow}$, is converted under
such a transformation to $n_{\alpha} - \hat{N}_{\alpha}$,
where $n_{\alpha}$ marks the total number of
electronic states in lead $\alpha$. Consequently,
$\hat{I}_{c, \alpha}$ transforms according to
\begin{equation}
\hat{I}_{c, \alpha} = -i e [ {\cal H}, \hat{N}_{\alpha} ]
                    \to -i e [ {\cal H},
                     n_{\alpha} - \hat{N}_{\alpha} ]
                     = -\hat{I}_{c, \alpha} .
\end{equation}
If the system begins its evolution from equilibrium,
i.e., the statistical averaging at time $t$ depends
solely on the Hamiltonian at previous times, then
the instantaneous charge current $I_{c, \alpha}(t)
\equiv \langle \hat{I}_{c, \alpha}(t) \rangle =
-\langle \hat{I}_{c, \alpha}(t) \rangle$ must
necessarily be zero.

The above argumentation is quite general, making no
reference to the microscopic details of ${\cal H}$,
nor to the temperature $T$. Its usefulness lies
in revealing the necessary (but not sufficient)
condition for a finite instantaneous charge current
to flow: Either the Hamiltonian is not permanently
invariant under the particle-hole transformation
indicated above, or the statistical averaging is
not determined by the Hamiltonian alone (as is the
case for a finite voltage bias). Note that this
symmetry bears no information on the spin current,
as the latter is invariant under the particle-hole
transformation specified above.

\subsection{Particle-hole symmetry that interchanges
            the two leads}
\label{sub-sec:PH-2}

An equivalent statement can be made about the symmetrized
spin current $I_s(t) = \langle \hat{I}_s(t) \rangle$
in case of a particle-hole symmetry that simultaneously
interchanges the two leads. Indeed, let us now assume
that ${\cal H}$ is invariant under a transformation
where particles on each lead are converted to
opposite-spin holes on the {\em opposite} lead (i.e.,
$c^{\dagger}_{\alpha, k, \sigma} \to e^{i \varphi_{\alpha\sigma}}
c_{\bar{\alpha}, -k, \bar{\sigma}}$, where
$\bar{\alpha}$ is the lead index opposite to $\alpha$).
Under such a transformation $\hat{I}_{c, \alpha}$
is converted to $-\hat{I}_{c, \bar{\alpha}}$, while
$\hat{I}_s$ is transformed to $-\hat{I}_s$. Hence, the
instantaneous spin current $I_s(t)$ must necessarily
be zero whenever evolution begins from thermal
equilibrium. By contrast, no general statement
can be made about the charge current in this
case, apart from the obvious identity
$I_{c, \alpha}(t) = -I_{c, \bar{\alpha}}(t)$.

\subsection{Application to the Kondo Hamiltonian}
\label{sub-sec:Kondo}

Our discussion thus far was quite general. We now
apply the symmetry arguments presented above to
the Kondo Hamiltonian of Eq.~(\ref{Full_H}). It
is easy to verify that Eq.~(\ref{Full_H}) is
invariant under the particle-hole transformation
\begin{equation}
\psi^{\dagger}_{\alpha \uparrow}(x)
      \to \psi_{\alpha \downarrow}(x) ,
\;\;\;\;\;
\psi^{\dagger}_{\alpha \downarrow}(x)
      \to -\psi_{\alpha \uparrow}(x)
\label{PH-K1}
\end{equation}
(corresponding to $\psi^{\dagger}_{\alpha,k,\sigma}
\to \pm \psi_{\alpha, -k, \bar{\sigma}}$),
regardless of the local field $H$ and the
Kondo couplings $J^{\alpha \beta}_{\lambda}
= J^{\beta \alpha}_{\lambda}$. Hence, the
instantaneous charge current for tunneling through
a Kondo impurity is strictly zero in the absence
of a voltage bias, as follows from the general
discussion of subsection~\ref{sub-sec:PH-1}. In
particular, no charge can be pumped through the
system unless a finite amplitude for potential
scattering is introduced into the Hamiltonian.
Although the description of real quantum dots
typically requires the inclusion of a
potential-scattering term, the latter can be made
negligibly small by operating the device deep in
the Kondo regime. In this manner charge transport
can be excluded.

Similarly, it is straightforward to confirm that
the Hamiltonian of Eq.~(\ref{Full_H}) is invariant
under the combined transformation
\begin{equation}
\psi^{\dagger}_{\alpha \uparrow}(x)
      \to \psi_{\bar{\alpha} \downarrow}(x) ,
\;\;\;\;\;
\psi^{\dagger}_{\alpha \downarrow}(x)
      \to -\psi_{\bar{\alpha} \uparrow}(x)
\label{PH-K2}
\end{equation}
(corresponding to $\psi^{\dagger}_{\alpha,k,\sigma}
\to \pm \psi_{\bar{\alpha}, -k, \bar{\sigma}}$),
provided the intra-lead exchange couplings obey
$J^{LL}_{\lambda} = J^{RR}_{\lambda}$. Thus,
the instantaneous spin current is strictly zero
if the spin couples equally to the two leads,
as follows from the general discussion of
sub-section~\ref{sub-sec:PH-2}. Spin pumping
therefore requires asymmetric coupling to the
two leads at least in some stretches of time.

It is instructive to re-derive these results based
on the symmetries of the transformed Hamiltonian
${\cal H}' + {\cal H}_{\rm int}$, which serves primarily
as a check for the correctness of Eqs.~(\ref{H'})
and (\ref{Hint}). Other than the free kinetic-energy
term, the flavor field $\psi_{f}$ enters both
${\cal H}'$ and ${\cal H}_{\rm int}$ only in the
form of $\hat{\chi}^{-}_{f}$, which is invariant
under the particle-hole transformation
\begin{equation}
\psi^{\dagger}_{f, k} \to -\psi_{f, -k} .
\label{ph-f}
\end{equation}
Note that the latter transformation is
restricted to the flavor sector. Consequently,
${\cal H}' + {\cal H}_{\rm int}$ is invariant
under the transformation of Eq.~(\ref{ph-f}),
while the charge-current operator, being proportional
to $\hat{\chi}^{+}_{f}$, transforms according to
$\hat{I}'_c \to -\hat{I}'_c$ [see Eq.~(\ref{I_c-int})].
This in turn demands that $I_c(t)$ be zero in the
absence of a voltage bias, in agreement with the
general symmetry considerations of Eq.~(\ref{PH-K1}).

Similarly, when $J_{\lambda}^{LL} = J_{\lambda}^{RR}$
the couplings $J_{sf}^{-}$ and $J_{z}^{-}$
drop from the transformed Hamiltonian
${\cal H}' + {\cal H}_{\rm int}$, which now
depends on the field $\psi_{sf}$ either through
the free kinetic-energy term, or in the form of
$\hat{\chi}^{+}_{sf}$. As a result the transformed
Hamiltonian is invariant under the spin-flavor
particle-hole transformation
\begin{equation}
\psi^{\dagger}_{sf, k} \to \psi_{sf, -k} ,
\end{equation}
while the spin-current operator, being proportional to
$\hat{\chi}^{-}_{sf}$, acquires an extra minus sign:
$\hat{I}'_s \to -\hat{I}'_s$ [see Eq.~(\ref{I_s-int})].
This in turn implies that the instantaneous spin
current $I_s(t)$ is strictly zero if the leads
couple equally to the spin, in agreement with
the symmetry considerations of Eq.~(\ref{PH-K2}).
Interestingly, $I_s(t)$ remains zero for symmetric
coupling also in the presence a finite bias, as
the latter couples solely to the flavor field. This
result, originally derived in Ref.~\onlinecite{SH98}
for nonequilibrium steady state, extends also to
time-dependent couplings and time-dependent bias.

\section{Pumped spin current}
\label{sec:spin-current}

Having established that the instantaneous charge
current vanishes for a generic Kondo model in the
absence of a voltage bias, we focus our attention
hereafter on the spin current. To this end, we
evaluate $I_s(t) = \langle \hat{I}_s(t) \rangle$
exactly on the Toulouse manifold by summing all
orders of the perturbation theory in the couplings
$J^{-}_{f}(t), J^{\pm}_{sf}(t)$, and $H(t)$. We
show that a finite spin current can indeed be
pumped through the system by applying a nonzero
magnetic field, provided the spin couples
asymmetrically to the two leads. The calculation
proceeds in three steps. Using the Keldysh technique,
we first derive a formal expression for the
instantaneous spin current in terms of the Majorana
Green functions of Eqs.~(\ref{G>_def})--(\ref{G-def}).
This portion of the derivation makes no assumption on
the time-dependent couplings, apart from the restriction
to the Toulouse manifold and the exclusion of an applied
voltage bias. The resulting expression is recast in
turn in subsection~\ref{TD-scattering-matrix} in
terms of a time-dependent scattering matrix for the
Majorana fermions, defined in Eq.~(\ref{S-def}) below.
The latter scattering matrix reduces in equilibrium to
the Fourier transform (with respect to energy) of the
conventional single-particle scattering matrix. Finally,
a systematic gradient expansion is carried out in
subsection~\ref{Gradient-expansion} for the case of
slowly varying potentials, resulting in a Brouwer-type
formula for the adiabatically pumped spin current.

\subsection{General formulation}

We begin by formally deriving the instantaneous spin
current $I_s(t)$ using the Keldysh technique, for
general time-dependent couplings on the Toulouse
manifold. As is always the case with the Keldysh
approach, we assume that the perturbations
$J_{sf}^{\pm}$, $J^{-}_{f}$, and $H$ have been
switched on at some distant time in the past,
$t_0 \to -\infty$, prior to which the system was
in thermal equilibrium.

To set the stage for the Keldysh formalism, the
spin current $I_s(t)$ is first written as
\begin{equation}
I_{s}(t) = \frac{i}{2}
           \left [
                   J^{-}_{sf}(t) G^{<}_{a, sf +}(t, t)
                   -  J^{+}_{sf}(t)
                   G^{<}_{b, sf -}(t, t)
           \right ] ,
\label{I_s-1}
\end{equation}
where
\begin{eqnarray}
G^{<}_{a, sf +}(t, t') &=&
      \langle \hat{\chi}_{sf}^{+}(t') \hat{a}(t) \rangle ,
\\
G^{<}_{b, sf -}(t, t') &=&
      \langle \hat{\chi}_{sf}^{-}(t') \hat{b}(t) \rangle
\end{eqnarray}
[see Eq.~(\ref{I_s})]. Using standard diagrammatics,
each of the latter correlators is expressed in an
exact manner as
\begin{eqnarray}
G^{<}_{a, sf +}(t, t') &=& -i \int_{-\infty}^{\infty}\!
         J_{sf}^{+}(\tau)
         \left [
                  G^{<}_{ab}(t,\tau)
                  g^{a}_{sf,+,+}(\tau, t')
         \right.
\nonumber \\
&& \;\;\;\;\;\;\;\;\;
         \left .
                  + G_{ab}^{r}(t,\tau)
                    g^{<}_{sf,+,+}(\tau,t')
         \right ] \! d\tau ,
\label{G_a,sf+} \\
G^{<}_{b, sf -}(t, t') &=& -i \int_{-\infty}^{\infty}\!
         J_{sf}^{-}(\tau)
         \left [
                  G^{<}_{ba}(t,\tau)
                  g^{a}_{sf,-,-}(\tau, t')
         \right.
\nonumber \\
&& \;\;\;\;\;\;\;\;\;
         \left .
                  + G_{ba}^{r}(t,\tau)
                    g^{<}_{sf,-,-}(\tau,t')
         \right ] \! d\tau ,
\label{G_b,sf-}
\end{eqnarray}
where
\begin{equation}
g^{<}_{\nu, p, p'}(t, t') =
    \langle
            \hat{\chi}_{\nu}^{p'}(t') \hat{\chi}_{\nu}^{p}(t)
    \rangle_{0}
\label{g^<}
\end{equation}
and
\begin{equation}
g^{a}_{\nu, p, p'}(t, t') = i \theta(t' - t)
    \langle \{
            \hat{\chi}_{\nu}^{p}(t), \hat{\chi}_{\nu}^{p'}(t')
            \}
    \rangle_{0}
\label{g^a}
\end{equation}
are the unperturbed Green functions for the
local Majorana fields. Here $\nu = f, sf$ and
$p, p' = \pm 1$. The zero subscripts in
Eqs.~(\ref{g^<}) and (\ref{g^a}) come to indicate
that both the time evolution and statistical
averaging are taken with respect to the
unperturbed Hamiltonian, i.e., the free
kinetic-energy part of Eq.~(\ref{H'}).

In writing Eq.~(\ref{G_a,sf+}) and (\ref{G_b,sf-}),
we have used the fact that $g^{<}_{f,-,+}$ and
$g^{a}_{f,-,+}$ identically vanish as long as no
voltage bias is applied.~\cite{comment_on_bias}
Indeed, in the wide-band limit Eqs.~(\ref{g^<})
and (\ref{g^a}) take the explicit forms
\begin{equation}
g^{<}_{\nu, p, p'}(t, t') =
      2 \pi \rho_0 \delta_{p p'} {\cal F}(t-t')
\label{g^<-explicit}
\end{equation}
and
\begin{equation}
g^{a}_{\nu, p, p'}(t, t') =
      i \pi \rho_0 \delta_{p p'} \delta(t -t') ,
\label{g^a-explicit}
\end{equation}
where ${\cal F}(t)$ is the Fourier transform of
the Fermi function $f(\epsilon)$:
\begin{eqnarray}
{\cal F}(t) = \lim_{\eta \rightarrow 0^{+}}
              \int_{-\infty}^{\infty}
                   \frac{d\epsilon}{2\pi}\,
                   e^{-i\epsilon t}\,
                   e^{-|\epsilon| \eta}\, f(\epsilon) .
\label{F-of-t}
\end{eqnarray}
The limiting procedure used in Eq.~(\ref{F-of-t})
corresponds to regularizing the conduction-electron
density of states per unit length according to
$\rho(\epsilon) = \rho_0 e^{-|\epsilon|\eta}$,
and taking the wide-band limit $D = 1/\eta \to \infty$.
Equation~(\ref{g^a-explicit}) is slightly modified
for a finite bandwidth $D$,~\cite{comment_on_finite_D}
but remains proportional to $\delta_{p p'}$.
Inserting Eqs.~(\ref{g^<-explicit})
and (\ref{g^a-explicit}) into Eqs.~(\ref{G_a,sf+})
and (\ref{G_b,sf-}), and plugging the resulting
expressions into Eq.~(\ref{I_s-1}), one obtains
\begin{widetext}
\begin{equation}
I_s(t) = i\frac{\pi \rho_0}{2}
               J_{sf}^{+}(t) J_{sf}^{-}(t)
          \left [
                   G^{<}_{ab}(t, t) - G^{<}_{ba}(t, t)
          \right ]
        + \pi \rho_0
          \int_{-\infty}^{\infty}
          [
                   J_{sf}^{-}(t) G^{r}_{ab}(t, \tau)
                   J_{sf}^{+}(\tau)
                 - J_{sf}^{+}(t) G^{r}_{ba}(t, \tau)
                   J_{sf}^{-}(\tau)
          ] {\cal F}(\tau - t) d\tau .
\label{I_s-via-G}
\end{equation}

It is easy to see at this point that the instantaneous
spin current vanishes in the absence of an applied
magnetic field, as it physically should. Indeed,
setting $H = 0$ in Eq.~(\ref{H'}), the two Majorana
fermions $\hat{a}$ and $\hat{b}$ decouple within the
Hamiltonian ${\cal H}'$. As a result the Green functions
$G_{ab}$ and $G_{ba}$ identically vanish, as does
$I_s$. It is also apparent that $I_s$ is strictly zero
unless the impurity couples asymmetrically to the two
leads, in accordance with the general symmetry arguments
of Sec.~\ref{sub-sec:Kondo}. In fact, $I_s(t)$ vanishes
not only when $J_{\perp}^{LL} = J_{\perp}^{RR}$ but
also for $J_{\perp}^{LL} = -J_{\perp}^{RR}$, which
stems from yet another symmetry of the Toulouse-limit
Hamiltonian. Specifically, Eq.~(\ref{H'}) is invariant
for $J_{sf}^{+} = 0$ under the particle-hole transformation
$\psi_{sf, k} \to -\psi_{sf, -k}^{\dagger}$, while
$\hat{I}_s$ transforms according to $\hat{I}_s \to
-\hat{I}_s$. Consequently $I_s(t) = - I_s(t)$ must
necessarily vanish when $J_{\perp}^{LL} = -J_{\perp}^{RR}$.

\subsection{Time-dependent scattering matrix}
\label{TD-scattering-matrix}

Although formally exact, Eq.~(\ref{I_s-via-G}) requires
knowledge of the time-dependent Green functions $G_{ab}$
and $G_{ba}$, which are difficult to compute for a
general time-dependent setting. In order to implement
the adiabatic limit, it is useful to first recast
Eq.~(\ref{I_s-via-G}) in terms of a time-dependent
scattering matrix to be defined below. This goal requires
a sequence of steps, starting with expressing the lesser
Green functions $G_{ab}^{<}$ and $G_{ba}^{<}$ in terms
of the retarded and advanced Green functions. Since the
Hamiltonian of Eq.~(\ref{H'}) is quadratic, one has
the identities
\begin{eqnarray}
G^{<}_{ab}(t,t) &=&
      2\pi\rho_0\!\int_{-\infty}^{\infty}\! d\tau
                \!\int_{-\infty}^{\infty}\! d\tau'
                \biggl[
                        G^r_{aa}(t,\tau)\!
                        \biggl\{
                                 J^{-}_{sf}(\tau)
                                 J^{-}_{sf}(\tau')
                                 + J^{-}_{f}(\tau)
                                   J^{-}_{f}(\tau')
                        \biggl\}
                        G^a_{ab}(\tau',t)
\nonumber \\
&& \;\;\;\;\;\;\;\;\;\;\;\;\;\;\;\;\;\;\;\;
   \;\;\;\;\;\;\;\;\;\;\;\;\;
                     +\, G^r_{ab}(t,\tau) J^{+}_{sf}(\tau)
                         J^{+}_{sf}(\tau') G^a_{bb}(\tau',t)
                \biggl]
                {\cal F}(\tau - \tau') ,
\label{G<-as-Gr-ab} \\
G^{<}_{ba}(t,t) &=&
      2\pi\rho_0\!\int_{\-\infty}^{\infty}\! d\tau
                \!\int_{\-\infty}^{\infty}\! d\tau'
                \biggl[
                        G^r_{ba}(t,\tau)\!
                        \biggl\{
                                 J^{-}_{sf}(\tau)
                                 J^{-}_{sf}(\tau')
                                 + J^{-}_{f}(\tau)
                                   J^{-}_{f}(\tau')
                        \biggl\}
                        G^a_{aa}(\tau',t)
\nonumber \\
&& \;\;\;\;\;\;\;\;\;\;\;\;\;\;\;\;\;\;\;\;
   \;\;\;\;\;\;\;\;\;\;\;\;\;
                     +\, G^r_{bb}(t,\tau) J^{+}_{sf}(\tau)
                         J^{+}_{sf}(\tau') G^a_{ba}(\tau',t)
                 \biggl]
                 {\cal F}(\tau - \tau') .
\label{G<-as-Gr-ba}
\end{eqnarray}
Substituting Eqs.~(\ref{G<-as-Gr-ab}) and
(\ref{G<-as-Gr-ba}) into Eq.~(\ref{I_s-via-G}), it
is convenient to introduce the scattering $T$-matrix
associated with the Majorana fields
$\hat{\chi}^{\pm}_{sf}$ and $\hat{\chi}^{-}_{f}$,
\begin{eqnarray}
{\bf T}^{r,a}(t,t') = 2\pi \rho_0
        \left[
               \begin{array}{ccc}
                     J^{+}_{sf}(t) G^{r,a}_{bb}(t,t')
                                   J^{+}_{sf}(t') &
                     J^{+}_{sf}(t) G^{r,a}_{ba}(t,t')
                                   J^{-}_{sf}(t') &
                     J^{+}_{sf}(t) G^{r,a}_{ba}(t,t')
                                   J^{-}_{f}(t')  \\ \\
                     J^{-}_{sf}(t) G^{r,a}_{ab}(t,t')
                                   J^{+}_{sf}(t') &
                     J^{-}_{sf}(t) G^{r,a}_{aa}(t,t')
                                   J^{-}_{sf}(t') &
                     J^{-}_{sf}(t) G^{r,a}_{aa}(t,t')
                                   J^{-}_{f}(t') \\ \\
                     J^{-}_{f}(t) G^{r,a}_{ab}(t,t')
                                   J^{+}_{sf}(t') &
                     J^{-}_{f}(t) G^{r,a}_{aa}(t,t')
                                   J^{-}_{sf}(t') &
                     J^{-}_{f}(t) G^{r,a}_{aa}(t,t')
                                   J^{-}_{f}(t') \\ \\
               \end{array}
        \right] .
\label{T-matrix}
\end{eqnarray}
Here the row and column indices $i = 1, 2, 3$ are
identified with $(sf,+), (sf,-)$, and $(f,-)$,
respectively. In terms of the $T$-matrix specified
above, the spin current is written as
\begin{eqnarray}
I_s(t) &=& \frac{1}{4}
           \left [
                   \int_{-\infty}^{\infty}\!d\tau\!
                   \left \{
                          {\bf T}^r(t,\tau){\cal F}(\tau-t)
                          -{\cal F}(t-\tau){\bf T}^a(\tau,t)
                   \right \}
              + i\!\int_{-\infty}^{\infty}\!d\tau\!\!
                   \int_{-\infty}^{\infty}\!d\tau'\,
                          {\bf T}^r(t,\tau)
                          {\cal F}(\tau - \tau')
                          {\bf T}^a(\tau',t)
           \right ]_{(sf-,sf+)}
\nonumber \\
       &-& \frac{1}{4}
           \left [
                   \int_{-\infty}^{\infty}\!d\tau\!
                   \left \{
                          {\bf T}^r(t,\tau){\cal F}(\tau-t)
                          -{\cal F}(t-\tau){\bf T}^a(\tau,t)
                   \right \}
              + i\!\int_{-\infty}^{\infty}\!d\tau\!\!
                   \int_{-\infty}^{\infty}\!d\tau'\,
                          {\bf T}^r(t,\tau)
                          {\cal F}(\tau - \tau')
                          {\bf T}^a(\tau',t)
           \right ]_{(sf+,sf-)} .
\label{current-via-Tmatrix}
\end{eqnarray}
Finally, the time-dependent scattering matrix for the
Majorana fields $\hat{\chi}^{\pm}_{\nu}$ is defined as
\begin{eqnarray}
\label{S-matrix}
\tilde{\bf S}(t,t') = \delta(t-t'){\bf 1}
                    - i\, {\bf T}^r(t,t') ,
\label{S-def}
\\
\tilde{\bf S}^{\dagger}(t,t') = \delta (t-t') {\bf 1}
                              + i\, {\bf T}^a(t,t') ,
\label{S-dag-def}
\end{eqnarray}
allowing us to compactly rewrite
Eq.~(\ref{current-via-Tmatrix}) in the form
\begin{equation}
I_s(t) = \frac{1}{2}{\rm Im}
         \int_{-\infty}^{\infty}\!d\tau
         \int_{-\infty}^{\infty}\!d\tau'
             \left [
                     \tilde{{\bf S}}(t,\tau)
                     {\cal F}(\tau - \tau')
                     \tilde{{\bf S}}^{\dagger}(\tau',t')
             \right ]_{(sf+,sf-)} .
\label{current-via-S-matrix}
\end{equation}
\end{widetext}

A word is in order at this point about the
time-dependent scattering matrix of Eq.~(\ref{S-def}).
Physically, $\tilde{{\bf S}}(t, t')$ describes the
scattering of an incoming Majorana fermion at time
$t'$ to an outgoing Majorana fermion at time $t$.
It reduces in equilibrium to the Fourier transform
(with respect to energy) of the conventional
single-particle scattering matrix, and remains
an exclusive function of the time difference
$\Delta t = t - t'$ under general steady-state
conditions. Although this ceases to be the
case in the presence of time-varying fields,
$\tilde{{\bf S}}(t, t')$ continues to satisfy
the generalized unitarity relation
\begin{equation}
\int_{-\infty}^{\infty}\!d\tau\,
       \tilde{{\bf S}}(t,\tau)
       \tilde{{\bf S}}^{\dagger}(\tau,t')
       = \delta(t - t') {\bf 1} ,
\label{unitarity}
\end{equation}
to be utilized below.

\subsection{Gradient expansion and Brouwer-type formula}
\label{Gradient-expansion}

The main achievement of Eq.~(\ref{current-via-S-matrix})
is the expression of the instantaneous spin current
in terms of the time-dependent scattering matrix
$\tilde{{\bf S}}(t, t')$. For a general periodic
modulation of the couplings $J^{\pm}_{sf}$, $J_{f}^{-}$,
and $H$, the instantaneous spin current at time $t$
depends on the specifics of the pumping cycle. For
example, the history and rates at which parameters are
varied. This is not the case in the adiabatic limit,
where the only information needed to predict the
pumped spin per cycle is (i) the shape of the
pumping trajectory in parameter space, and (ii) the
equilibrium $S$-matrix along the trajectory. Similar
to adiabatic quantum pumping in noninteracting
systems, the adiabatic limit is approached when the
characteristic modulation frequency $\Omega$ obeys
$\Omega \ll \Gamma_a, \Gamma_b$ at each point along
the pumping trajectory. Here $\Gamma_a$ and $\Gamma_b$
are the energy scales defined in Eqs.~(\ref{Gamma_a})
and (\ref{Gamma_b}), respectively.

To substantiate these claims and devise a Brouwer-type
formula for adiabatic quantum spin pumping in the Kondo
regime, we resort to a systematic gradient expansion
of Eq.~(\ref{current-via-S-matrix}). To this end, we
first introduce the Wigner transform of the
time-dependent scattering matrix,
\begin{equation}
{\bf S}(\epsilon,T) =
       \int_{-\infty}^{\infty}\!d\tau\,
            e^{i\epsilon \tau}\,
            \tilde{{\bf S}}
            \left (
                    T + \frac{\tau}{2},
                    T - \frac{\tau}{2}
            \right ) .
\label{WignerT}
\end{equation}
Next we apply the well-developed machinery of the
Gradient expansion.~\cite{Haug-Jauho-book} For
example, the Wigner transform of the convolution
of two functions,
\begin{widetext}
\begin{equation}
[A \star B](\epsilon,T) =
            \int_{-\infty}^{\infty}\!d\tau
                   e^{i\epsilon \tau}\!
            \int_{-\infty}^{\infty}\!d\tau'
                   A \left(
                            T + \frac{\tau}{2},\tau'
                     \right )\!
                   B \left( \tau',
                            T - \frac{\tau}{2}
                     \right ) ,
\end{equation}
has the formally exact
representation~\cite{Haug-Jauho-book}
\begin{equation}
[A\star B](\epsilon,T) =
     e^{
             \frac{1}{2i}
             \left \{
                      \partial_T^A \partial_{\epsilon}^B
                    - \partial_T^B \partial_{\epsilon}^A
             \right\}
       }
     A(\epsilon,T) B(\epsilon,T)
     = A(\epsilon,T) B(\epsilon,T)
       + \frac{1}{2i}
       \left (
               \partial_T A \partial_{\epsilon} B
             - \partial_{\epsilon} A \partial_{T} B
       \right ) + \cdots .
\label{GE-of-convolution}
\end{equation}
Here $\partial^A$ and $\partial^B$ stand for
differential operators that act on $A(\epsilon, T)$
and $B(\epsilon, T)$, respectively. The usefulness of
Eq.~(\ref{GE-of-convolution}) comes into play when
the expansion on the right-hand side is controlled
by a small parameter. This is indeed the case in the
present context, where the double convolution of
Eq.~(\ref{current-via-S-matrix}) possesses an analogous
expansion in gradients of $S(\epsilon, T)$. Each
combined derivative $\partial_T \partial_{\epsilon}$ is
parametrically reduced for Eq.~(\ref{current-via-S-matrix})
by a factor of $\Omega/\bar{\Gamma}$, where $\bar{\Gamma}$
is some characteristic value of either $\Gamma_a$ or
$\Gamma_b$ in the relevant time interval. The scale
$\bar{\Gamma}$ is bounded from below by the minimum
of $\Gamma_a$ and $\Gamma_b$ along the pumping cycle,
a quantity denoted hereafter by $\Gamma$. Hence, for
$\Omega \ll \Gamma$ one can settle with linear order
in $\partial_T \partial_{\epsilon}$ to obtain
\begin{equation}
I_s(t) = {\rm Im}
   \int_{-\infty}^{\infty}
         \frac{d\epsilon}{4 \pi}\!
         \left [
                 \left \{
                         \!{\bf S}{\bf S}^{\dagger}
                         \!+\!\frac{1}{2i}
                         \left [
                                (\partial_t {\bf S})
                                (\partial_{\epsilon}
                                       {\bf S}^{\dagger})
                              - (\partial_{\epsilon}
                                       {\bf S})
                                (\partial_{t}
                                       {\bf S}^{\dagger})
                         \right ]
                 \right \} f(\epsilon)
               + \frac{1}{2i}
                 \left \{
                         {\bf S}(\partial_{t}
                                  {\bf S}^{\dagger})
                       - (\partial_t {\bf S})
                                  {\bf S}^{\dagger}\!
                 \right \}
                 \left (
                         -\frac{\partial f(\epsilon)}
                               {\partial \epsilon}
                 \right )\!
         \right ]_{(sf+,sf-)} .
\label{current-as-WT}
\end{equation}
\end{widetext}
All terms omitted in this expression are of order
$(\Omega/\Gamma)^2$ or higher, and thus can be
safely neglected.~\cite{comment_on_omega^2}

The term proportional to the Fermi function $f(\epsilon)$
in Eq.~(\ref{current-as-WT}) is purely diagonal to
order ${\cal O} \left ( \Omega/\Gamma \right )$, as
can be seen by expanding the unitarity relation of
Eq.~(\ref{unitarity}) to first order in time
gradients:
\begin{equation}
{\bf S}{\bf S}^{\dagger} + \frac{1}{2i}
       \left [
                (\partial_t {\bf S})
                (\partial_{\epsilon} {\bf S}^{\dagger})
              - (\partial_{\epsilon} {\bf S})
                (\partial_{t}{\bf S}^{\dagger})
       \right ]
       + {\cal O}
         \left [
                 (\Omega/\Gamma)^2
         \right ] = {\bf 1} .
\end{equation}
Since Eq.~(\ref{current-as-WT}) requires an off-diagonal
matrix element of the expression in the square brackets,
the instantaneous spin current reduces in the adiabatic
limit to
\begin{equation}
I_s(t) = {\rm Re}
         \int_{-\infty}^{\infty}
              \frac{d\epsilon}{8\pi}
              f'(\epsilon)
              \biggl [
                       {\bf S}
                       (\partial_{t} {\bf S}^{\dagger})
                     - (\partial_t {\bf S})
                       {\bf S}^{\dagger}
             \biggl ]_{(sf+,sf-)} .
\label{current-final}
\end{equation}
This expression can further be simplified by noting
that $S(\epsilon, t)$ is equal to leading order in
$\Omega/\Gamma$ to the instantaneous scattering
matrix, i.e., the equilibrium scattering matrix with
all system parameters $J^{\pm}_{sf}$, $J_{f}^{-}$, and
$H$ frozen at their instantaneous values at time $t$:
\begin{equation}
{\bf S}(\epsilon,t) =
       {\bf S}_{\rm eq}(\epsilon; J^{\pm}_{sf}(t),
                                  J^{-}_f(t),H(t))
       + {\cal O} \left ( \Omega/\Gamma \right ) .
\label{instantaneous}
\end{equation}
Consequently, one can substitute ${\bf S}_{\rm eq}$
in for ${\bf S}$ in Eq.~(\ref{current-final}).
Lastly, one can exploit the unitarity of the
equilibrium $S$-matrix, ${\bf S}_{\rm eq}
{\bf S}^{\dagger}_{\rm eq} = {\bf 1}$, to
replace $(\partial_t {\bf S}_{\rm eq})
{\bf S}^{\dagger}_{\rm eq}$ with $-{\bf S}_{\rm eq}
(\partial_t {\bf S}^{\dagger}_{\rm eq})$ in
Eq.~(\ref{current-final}). This yields the final
expression for the spin current,
\begin{equation}
I_s(t) = {\rm Re}
    \left \{
             \int_{-\infty}^{\infty}
                  \frac{d\epsilon}{4\pi}
                  f'(\epsilon)
                  \left [
                         {\bf S}
                         (\partial_{t} {\bf S}^{\dagger})
                  \right ]_{(sf+,sf-)}
    \right \} .
\label{current-final-final}
\end{equation}
Here and in the remainder of the paper the symbol
${\bf S}$ is used as a shorthand for the
instantaneous scattering matrix ${\bf S}_{\rm eq}
(\epsilon; J^{\pm}_{sf}(t), J^{-}_f(t),H(t))$.

Equation~(\ref{current-final-final}) is exact in the
adiabatic limit, $\Omega \to 0$. Its derivation was
based on a systematic truncation of higher order
terms in $\Omega$, controlled by the expansion
parameter $\Omega/\Gamma$. It therefore encompasses
all pumping trajectories and all coupling regimes,
whether weak or strong. This should be contrasted with
the commonly used linear-response theory, which is
restricted, strictly speaking, to weak coupling only.

In the following we shall consider examples of pumping
cycles where two system parameters, generically termed
$X_1$ and $X_2$, are varied slowly and periodically
in time along a certain closed trajectory ${\cal C}$
in parameter space. The quantity of interest in this
case is the total magnetization in the $z$ direction,
or spin, transferred from left to right in a single
pumping cycle. The latter quantity is defined as
\begin{eqnarray}
\langle S \rangle = \oint_{{\cal C}} \!I_s(t)\,dt ,
\end{eqnarray}
where $I_s(t)$ is the instantaneous spin current.
Using Eq.~(\ref{current-final-final}) one can
express $\langle S \rangle$ as a line integral
along the contour ${\cal C}$,
\begin{equation}
\langle S \rangle = {\rm Re}
    \left \{
             \int_{-\infty}^{\infty}
                  \frac{d\epsilon}{4\pi}
                  f'(\epsilon)
                  \oint_{{\cal C}}
                  \left [
                         {\bf S}
                         \underline{\nabla}
                              {\bf S}^{\dagger}
                  \right ]_{(sf+,sf-)}
                  \cdot d\underline{X}
    \right \} .
\end{equation}
This expression applies to the variation of any
number of system parameters $X_1, \cdots, X_N$. In
the particular case where $N = 2$, one can make use of
Green's theorem to express the spin pumped per cycle
as a geometric property of the Majorana-fermion
scattering matrix, analogous to Brouwer's formula
for noninteracting systems. Explicitly,
$\langle S \rangle$ assumes the form
\begin{eqnarray}
\langle S \rangle &=& {\rm Re}
          \int_{-\infty}^{\infty}
               \frac{d\epsilon}{4\pi} f'(\epsilon)
               \int_{{\cal A}} dX_1 dX_2
\nonumber \\
           &&\times \left [
                       \partial_{X_1}\!{\bf S}\,
                       \partial_{X_2}{\bf S}^{\dagger}
                     - \partial_{X_2}{\bf S}\,
                       \partial_{X_1}\!{\bf S}^{\dagger}
                    \right ]_{(sf+,sf-)} ,
\label{spin-as-area-integral}
\end{eqnarray}
where ${\cal A}$ is the (oriented) area in parameter
space enclosed by the contour ${\cal C}$.

Equation~(\ref{spin-as-area-integral}) is the central
result of our study. We devote the remainder of the
paper to analyzing its implications for a particular
class of pumping trajectories defined below.

\section{Applications}
\label{sec:results}

\begin{figure}
\centerline{
\includegraphics[width=70mm]{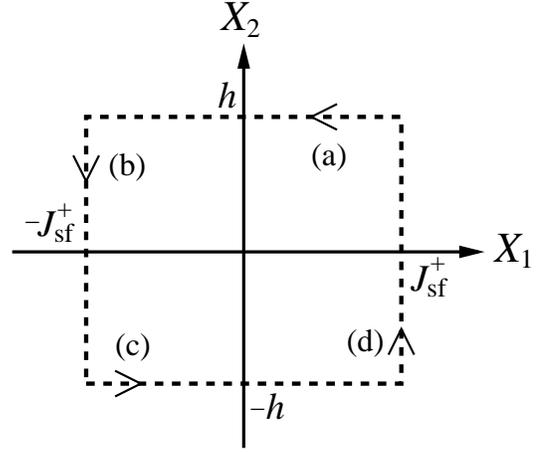}
}\vspace{0cm}
\caption{The pumping cycle under consideration in
         Sec.~\ref{sec:results}. The first pumping
         parameter, $X_1$, controls the Kondo couplings
         $J_{sf}^{-}$ and $J_{f}^{-}$, which vary
         according to $J_{sf}^{-} = X_1$ and
         $J_{f}^{-} = \sqrt{(J_{sf}^{+})^2 - X_1^2}$.
         The third Kondo coupling, $J_{sf}^{+}$, is held
         fixed throughout the cycle, along with $T_K$.
         The second pumping parameter, $X_2$, controls
         the Zeeman splitting $\mu_B g_i H$, which
         varies according to $\mu_B g_i H = X_2$.}
\label{Fig:fig2}
\end{figure}

We conclude our analysis by applying the formula
derived above to study a particularly simple class
of pumping cycles where one parameter, $X_1$,
controls the transverse Kondo couplings, and the
other parameter, $X_2$, controls the applied magnetic
field. Based on our Toulouse-limit calculations we
will show that such a cycle can be used to realize a
pure quantized spin pump, namely, {\em quantized}
spin pumping without any charge transport.

To make contact with realistic systems such as
quantum dots, we impose hereafter the condition
$J^{LL}_{\perp} J^{RR}_{\perp} = (J^{LR}_{\perp})^2$,
corresponding to $(J_{sf}^{-})^2 + (J_{f}^{-})^2 =
(J_{sf}^{+})^2$. As mentioned above, this condition
best describes the strong-coupling physics of the
Anderson impurity model, where a single Kondo scale
$\Gamma_a = \Gamma_b \equiv T_K$ emerges. Keeping
$J_{sf}^{+}$, and thus $T_K$, fixed, we parameterize
$J_{sf}^{-}$, $J_{f}^{-}$, and $\mu_B g_i H$
according to
\begin{eqnarray}
J_{sf}^{-} &=& X_1 ,
\label{J_sf-via-X1}
\\
J_{f}^{-} &=& \sqrt{ (J_{sf}^{+})^2 - X_1^2 } ,
\label{J_f-via-X1}
\end{eqnarray}
and
\begin{equation}
\mu_B g_i H = X_2 .
\end{equation}
In terms of the original Kondo couplings to
the two leads, Eqs.~(\ref{J_sf-via-X1}) and
(\ref{J_f-via-X1}) translate to
\begin{eqnarray}
&& J_{\perp}^{LL/RR} = \sqrt{2 \pi a}
          \left (
                  J_{sf}^{+} \pm X_1
          \right ) ,
\label{J_LL-via-X1}
\\
&& J_{\perp}^{LR} = \sqrt{2 \pi a}
          \sqrt{ (J_{sf}^{+})^2 - X_1^2 } .
\label{J_LR-via-X1}
\end{eqnarray}

The pumping cycle under consideration is depicted
schematically in Fig.~\ref{Fig:fig2}. It consists
of four segments, two in which $X_1$ is tuned from
$\pm J_{sf}^{+}$ to $\mp J_{sf}^{+}$ while $X_2$ is
kept fixed [lines (a) and (c)], and two in which
$X_2$ is tuned from $\pm h$ to $\mp h$ while $X_1$
is held fixed [lines (b) and (d)]. The cycle ${\cal C}$
thus consists of periodic opening/closing of the
transverse couplings to the left/right leads, followed
by inversion of the applied magnetic field at points
where spin-flip scattering is restricted to one
lead only.
The analogous cycle for real quantum dots comprises
of periodic opening/closing of the tunneling rates
to the left/right leads, followed by inversion of
the applied magnetic field at points where tunneling
is restricted to one lead only.

\begin{figure}[tb]
\centerline{
\includegraphics[width=70mm]{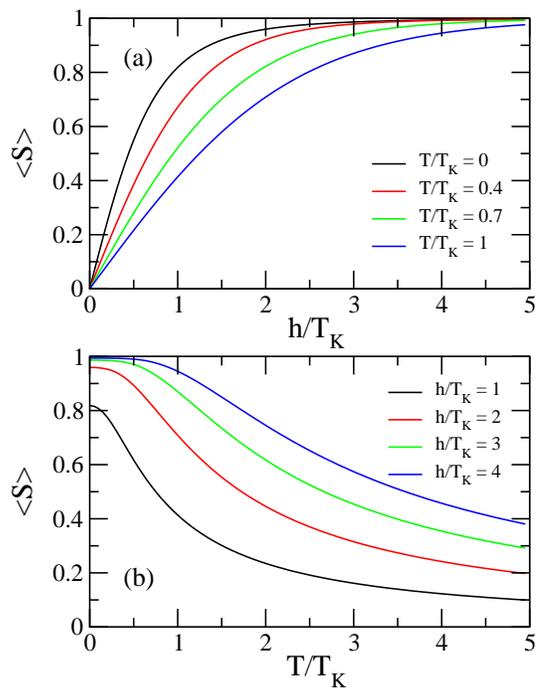}
}\vspace{0.3cm}
\caption{The spin pumped per cycle, $\langle S \rangle$,
         in units of $\hbar$. (a) Plotted as a function
         of $h/T_K$ for different $T$; (b) Plotted as a
         function of $T/T_K$ for different $h$. For
         $T < T_K$, the spin pumped per cycle rapidly
         approaches $\hbar$ with increasing $h$.
         Explicitly, $\langle S \rangle$ exceeds
         $0.9 \hbar$ for all $T \leq 0.45 T_K$ when
         $h = 2 T_K$.}
\label{Fig:fig3}
\end{figure}

Combining Eq.~(\ref{spin-as-area-integral}) for
$\langle S \rangle$ with Eqs.~(\ref{S-def}),
(\ref{S-dag-def}), (\ref{T-matrix}), and (\ref{G^ra})
for the instantaneous $S$-matrix, one obtains after
some straightforward but tedious algebra
\begin{eqnarray}
\langle S \rangle &=& -\frac{2 \hbar}{\pi}
     \int_{-\infty}^{\infty}
          d\epsilon f'(\epsilon)
          \left [
                  {\rm Re}
                  \left \{
                           \frac{h\,T_K}
                                {h^2 +
                                 (T_K - i\epsilon)^2}
                  \right \}
          \right.
\nonumber \\
&& \;\;\;\;\;\;\;\;\;\;\;\;\;\;\;\;
   \;\;\;\;\;\;\;\;\;\;\;\;\;
          \left.
                  + \arctan
                    \left (
                            \frac{h + \epsilon}{T_K}
                    \right )
          \right ] .
\label{example}
\end{eqnarray}
Here we have restored $\hbar$ for proper units of
$\langle S \rangle$. Representative plots of
$\langle S \rangle$ as a function of both $h$ and
$T$ are shown in Fig.~\ref{Fig:fig3}. As expected
of the Kondo regime, $\langle S \rangle$ is an
exclusive function of the rescaled parameters
$h/T_K$ and $T/T_K$. In particular, at $T = 0$
one finds
\begin{eqnarray}
\langle S \rangle = \frac{2 \hbar}{\pi}
          \left [
                  \left (
                          \frac{h}{T_K}
                          + \frac{T_K}{h}
                  \right )^{-1}\!
                  + \;\arctan
                    \left (
                            \frac{h}{T_K}
                    \right )
          \right ] ,
\end{eqnarray}
which has the formal expansion $\langle S \rangle
/\hbar = 1 - {\cal O} \left [ (T_K/h)^3 \right ]$.
Hence, the pumped spin per cycle is closely
quantized to $\hbar$ when the magnetic field $H$
performs a large enough excursion along the pumping
cycle. The effect of a temperature is to reduce the
spin pumped per cycle. However, $\langle S \rangle$
remains closely quantized to $\hbar$ when
$h \gg T_K, T$. Importantly, when $T \ll T_K$, it
suffices that $h$ will only moderately exceed $T_K$
in order for $\langle S \rangle$ to closely approach
$\hbar$. For example, at $T = 0$ the pumped spin
per cycle is equal to $0.82 \hbar$ ($0.96 \hbar$)
by the time $h = T_K$ ($h = 2T_K$).

The above results were derived at the Toulouse limit,
which does not correspond to any realistic parameters.
It is therefore pertinent to question the relevance
of these results to actual quantum dots. Since any
exact solution can be used to extract universal
low-energy properties of the Kondo effect, we expect
Eq.~(\ref{example}) to be {\em quantitatively} correct
when $T, h \ll T_K$. Equation~(\ref{example}) should
remain qualitatively correct as one of the parameters
$T$ or $h$ becomes comparable to $T_K$, though
quantitative deviations are expected. Still, since
$\langle S \rangle$ approaches $\hbar$ quite rapidly
with increasing $h$ (essentially by $h \sim T_K$),
and since the departure from strong coupling
is only logarithmically slow in $h$, we expect
$\langle S \rangle$ to remain nearly quantized
in real quantum dots provided $T \ll T_K$. This
picture is further supported by a naive application
of Brouwer's formula using the exact $T = 0$
single-particle scattering matrix,~\cite{Langreth66}
and by slave-boson mean-field theory of the
corresponding Anderson model.~\cite{Aono04} The
Toulouse limit fails, however, to describe the
weak-coupling regime, as certain bare couplings
are required to be large. In particular,
Eq.~(\ref{example}) should neither be quantitatively
nor qualitatively correct when $T_K \ll T$.

A simple interpretation of Eq.~(\ref{example}) follows
from the observation that the ground state of the Kondo
model is that of a local Fermi liquid. Only resonant
elastic scattering takes place at the Fermi level when
$T = 0$, as reflected in the Abrikosov-Shul resonance.
The latter resonance is pinned to the Fermi energy when
$H = 0$, and is split by an applied magnetic field.
This basic phenomenology can be mimicked by a simple
noninteracting resonant-level model,
\begin{eqnarray}
{\cal H}_{\rm RLM} &=&\!\sum_{\alpha = L,R}
             \sum_{k, \sigma} \epsilon_k
                  \psi^{\dagger}_{k \alpha \sigma}
                  \psi_{k \alpha \sigma}
                  - \mu_B g_i H
                  ( d_{\uparrow}^{\dagger}
                    d_{\uparrow} -
                    d_{\downarrow}^{\dagger}
                    d_{\downarrow} )
\nonumber \\
         &&+\!    \sum_{k, \alpha, \sigma}
                       V_{\alpha}\,
                       \{ \psi^{\dagger}_{k \alpha \sigma}
                          d_{\sigma} + {\rm H.c.} \} ,
\label{RLM}
\end{eqnarray}
which is studied below. Here
$\psi^{\dagger}_{k \alpha \sigma}$ creates an
electron with wave number $k$ and spin projection
$\sigma$ on lead $\alpha$ ($\alpha = L, R$), while
$d^{\dagger}_{\sigma}$ creates a localized electron
on the level.

Allowing for slow periodic modulation of $H$ and
$V_{\alpha}$ in Eq.~(\ref{RLM}), we extract the
adiabatically pumped spin and charge along a closed
pumping cycle analogous to the one shown in
Fig.~\ref{Fig:fig2}. For a generic trajectory
${\cal C}$ in the parameter space $(X_1, X_2)$
defined below, the adiabatically pumped spin and
charge are given for ${\cal H}_{\rm RLM}$ by the
standard Brouwer formula
\begin{eqnarray}
\langle S \rangle &=& \hbar \sum_{\sigma} \sigma
          \int_{-\infty}^{\infty}
                 \frac{d\epsilon}{4\pi i}
                 f'(\epsilon)
          \int_{{\cal A}} dX_1 dX_2
\nonumber \\
       && \times
          \left [
               \partial_{X_1}\!{\bf S}_{\sigma}\,
               \partial_{X_2}{\bf S}_{\sigma}^{\dagger}
             - \partial_{X_2}{\bf S}_{\sigma}\,
               \partial_{X_1}\!{\bf S}_{\sigma}^{\dagger}
          \right ]_{LL}
\label{S-RLM}
\end{eqnarray}
and
\begin{eqnarray}
\langle Q \rangle &=& -e \sum_{\sigma}
          \int_{-\infty}^{\infty}
                 \frac{d\epsilon}{2\pi i}
                 f'(\epsilon)
          \int_{{\cal A}} dX_1 dX_2
\nonumber \\
       && \times
          \left [
               \partial_{X_1}\!{\bf S}_{\sigma}\,
               \partial_{X_2}{\bf S}_{\sigma}^{\dagger}
             - \partial_{X_2}{\bf S}_{\sigma}\,
               \partial_{X_1}\!{\bf S}_{\sigma}^{\dagger}
          \right ]_{LL} .
\end{eqnarray}
Here $\sigma = \uparrow, \downarrow$ and $\sigma = \pm 1$
are used interchangeably to label the spin projection.
The domain of integration, ${\cal A}$, is the (oriented)
area in parameter space enclosed by the contour
${\cal C}$. The instantaneous $S$-matrix pertaining
to ${\cal H}_{\rm RLM}$ is written in the $L$-$R$
basis as
\begin{equation}
{\bf S}_{\sigma}(\epsilon) \!=\!
     \left[
            \begin{array}{cc}
                  1 - 2i\Gamma_L\,G^r_{\sigma}(\epsilon) &
                 -2 i \sqrt{\Gamma_L \Gamma_R}
                         G^r_{\sigma}(\epsilon) \\ \\
                 -2 i \sqrt{\Gamma_L \Gamma_R}
                         G^r_{\sigma}(\epsilon) &
                  1 - 2i\Gamma_R\,G^r_{\sigma}(\epsilon)
          \end{array}
          \right] ,
\end{equation}
where
\begin{equation}
G^{r}_{\sigma}(\epsilon) =
  \frac{1}{\epsilon - \sigma \mu_B g_i H + i \Gamma_{+}}
\label{G_function_RLM}
\end{equation}
is the associated dot Green function. Here
$\Gamma_{+} = \Gamma_{L} + \Gamma_{R}$ with
$\Gamma_{\alpha} = \pi \rho_0 V_{\alpha}^2$
is the resonance width, which plays the role
of the Kondo temperature in the Kondo model.

By analogy with the cycle of Fig.~\ref{Fig:fig2}, we
vary the two pumping parameters $X_1 = \Gamma_{L} -
\Gamma_{R}$ and $X_2 = \mu_B g_i H$ while $\Gamma_{+}$
is held fixed. As before, the cycle is composed
of four segments, two in which $X_1$ is tuned from
$\pm \Gamma_{+}$ to $\mp \Gamma_{+}$ while $X_2$ is
kept fixed, and two in which $X_2$ is tuned from
$\pm h$ to $\mp h$ while $X_1$ is held fixed. Using
Eqs.~(\ref{S-RLM})--(\ref{G_function_RLM}) for this
cycle one obtains $\langle Q \rangle = 0$ and
\begin{eqnarray}
\langle S \rangle &=& -\frac{2 \hbar}{\pi}
     \int_{-\infty}^{\infty}
          d\epsilon f'(\epsilon)
          \left [
                  {\rm Re}
                  \left \{
                           \frac{h\,\Gamma_{+}}
                                {h^2 + (\Gamma_{+}
                                        - i\epsilon)^2}
                  \right \}
          \right.
\nonumber \\
&& \;\;\;\;\;\;\;\;\;\;\;\;\;\;\;\;
   \;\;\;\;\;\;\;\;\;\;\;\;\;
          \left.
                  + \arctan
                    \left (
                            \frac{h + \epsilon}{\Gamma_{+}}
                    \right )
          \right ] .
\label{<S>-RLM}
\end{eqnarray}
Both results are identical to those obtain at the
Toulouse limit, provided $\Gamma_{+}$ is identified
with $T_K$. Thus, the physical picture underlying
Eq.~(\ref{example}) is consistent with that of
simple resonant elastic scattering, where a
single resonance is symmetrically split about
the Fermi energy by an applied magnetic field.

From a theoretical standpoint it is clear that
one can realize a quantized spin pump using either
a quantum dot in the Kondo regime or a Zeeman-split
single-particle resonance that is tuned to the Fermi
energy. However, practical considerations make the
Kondo-dot scenario a more promising candidate for the
realization of such a device. Indeed, modulation of the
couplings to the two leads is typically accompanied
in real devices by a capacitive shift of the dot
level. In case of a simple resonance, the induced
modulation of the dot level will generally produce
a finite charge current, and is likely to spoil
the quantization of the pumped spin. The Kondo-dot
scenario is immune to such fluctuations, as these
produce only a tiny shift of the Abrikosov-Shul
resonance. Indeed, as discussed in
Sec.~\ref{sec:symmetries}, charge transport is
strictly forbidden as long as the Coulomb-blockaded
dot can be described in terms of a pure Kondo
Hamiltonian having no potential scattering. Although
a realistic description of quantum dots generally
requires the inclusion of potential scattering, the
latter term can be made negligibly small by operating
the device deep in the Kondo regime. In this manner
charge transport can be excluded.

\section{Conclusions}
\label{sec:discussion}

In this paper we have presented an exact analysis
of adiabatic quantum pumping through a quantum dot
in the Kondo regime. It follows from general symmetry
arguments that the instantaneous charge current is
strictly zero in the absence of potential scattering
and for zero voltage bias. A similar statement applies
to the symmetrized spin current either in the absence
of an applied magnetic field or for symmetric coupling
to the leads. Pumping of a spin current therefore
requires both a finite magnetic field and for
left-right symmetry to be simultaneously broken. Both
conditions are readily met in practical devices,
making ultrasmall quantum dots a natural candidate
for the realization of a spin battery.

To quantify this statement, we have computed the
pumped spin current exactly at the Toulouse limit.
Exploiting the mapping onto a quadratic Hamiltonian
and performing a controlled expansion in the
small parameter $\Omega/T_K$ ($\Omega$ being the
characteristic modulation frequency, $T_K$ is the
Kondo temperature), we have expressed the pumped spin
per cycle as a geometric property of the scattering
matrix associated with three flavors of Majorana
fermions, which are the effective degrees of
freedom at the Toulouse limit. In particular,
employing the coupling to the leads as one pumping
parameter and the applied magnetic field as another,
we have shown that one can devise pumping cycles
that realize a pure quantized spin pump. Namely,
a device for which the average spin pumped per
cycle is closely equal to $\hbar$, but where no
accompanying charge current is produced. We
expect the pumped spin per cycle to remain
nearly quantized in real quantum dots provided
that one operates at $T \ll T_K$.

There have been by now a number of different proposals
in the literature for the realization of spin pumps,
employing diverse setups such as chaoticity in quantum
dots,~\cite{MCM02} ferromagnetic leads,~\cite{Zheng03}
spin-orbit interactions,~\cite{GTF03, SB03} classical
turnstile cycles,~\cite{BF05} one-dimensional
Luttinger-liquid physics,~\cite{SC01-03} and finally
the Kondo effect in quantum dots.~\cite{Aono04}
While all these proposals reported schemes to
realize a pure adiabatic spin pump along specific
cycles, the quantization of the spin pumped per
cycle has been shown to be the case only in the
classical turnstile setup~\cite{BF05} and for a
Luttinger liquid.~\cite{SC01-03} In contrast
to Ref.~\onlinecite{BF05}, the pumping scheme
investigated in this paper offers an interesting
possibility to realize a {\em coherent} quantized
spin pump, in which the absence of charge current
is essentially warranted along all possible cycles
(including beyond the adiabatic limit).

The quantization of the pumped spin per cycle
reported in this paper is subject to small deviations
as the temperature $T$ becomes of order $T_K$, or as
the magnetic-field excursion is altered. Moreover, it
applies only to the {\em average} spin pumped per cycle.
In order to better characterize such a quantum pump,
a detailed study of its noise properties (and full
counting statistics) is desirable. A study of the
statistical properties of the Kondo pump is a challenge
left for future work.

\section*{Acknowledgments}

A. Silva would like to thank N. Andrei, R. Fazio,
Y. Gefen, and in particular Y. Oreg and E. Sela
for many instructive discussions on the subject
of quantum pumping. A. Schiller is grateful to
S. Hershfield for an earlier collaboration on the
Toulouse limit. A. Schiller was supported in part
by the Center of Excellence Program of the Israel
Science Foundation.  Part of this work was performed
while A. Silva was visiting the Braun Submicron
Center at the Weizmann Institute of Science,
supported by EU grant RITA-CT-2003-506095.


\begin{thebibliography}{9}
\bibitem{comment_on_finite_D}
        Explicitly, the delta function $\delta(t - t')$
        in Eq.~(\ref{g^a-explicit}) is replaced with a
        Lorentzian of half-width $\eta = 1/D$.
\bibitem{comment_on_omega^2}
        Although our expansion is controlled by the
        small parameter $\Omega/\Gamma$, and is thus
        exact for $\Omega \to 0$, it does not provide
        us with a universal criterion just how small
        must $\Omega/\Gamma$ be for the higher order
        terms to become negligible. This will generally
        depend on details of the cycle of interest.
\end{thebibliography}
\end{document}